\documentclass[12pt]{article}
\usepackage{amssymb}
\usepackage{epsfig}
\usepackage{cite}

\def\be{\begin{equation}}
\def\ee{\end{equation}}
\def\bea{\begin{eqnarray}}
\def\eea{\end{eqnarray}}
\def\beb{\begin{eqnarray*}}
\def\eeb{\end{eqnarray*}}
\def\pat{\partial}

\begin{document}
\makeatletter
\def\fmslash{\@ifnextchar[{\fmsl@sh}{\fmsl@sh[0mu]}}
\def\fmsl@sh[#1]#2{%
  \mathchoice
    {\@fmsl@sh\displaystyle{#1}{#2}}%
    {\@fmsl@sh\textstyle{#1}{#2}}%
    {\@fmsl@sh\scriptstyle{#1}{#2}}%
    {\@fmsl@sh\scriptscriptstyle{#1}{#2}}}
\def\@fmsl@sh#1#2#3{\m@th\ooalign{$\hfil#1\mkern#2/\hfil$\crcr$#1#3$}}
\makeatother

\thispagestyle{empty}
\begin{titlepage}
        
\hspace*{\fill}{{\normalsize \begin{tabular}{l}
                              %{\sf physics/0602105n}\\
                              {\sf UWThPh-2006-2}\\
                              \end{tabular} }}

\boldmath
\begin{center}
  {\Large {\bf Non-Commutative Geometry \& Physics}\footnote{Seminar talks given at the 
  Universities Ivano-Frankivsk and Kamenets-Podolsk (Ukraine), April 2005.}}
\end{center}
\unboldmath

\setcounter{footnote}{0}
\renewcommand{\thefootnote}{\arabic{footnote}}
\begin{center}
                                                                                                                                         
{{{\bf
Michael~Wohlgenannt
}}}
                                                                                                                                         
\end{center}
\vskip 1em
\begin{center}
Institute for Theoretical Physics, \\
University of Vienna,\\
  Boltzmanngasse 5, 1090 Wien, Austria\\
  Email: michael.wohlgenannt@univie.ac.at                                                                                                              
\end{center}

\vspace{\fill}
                                                                                                                                         
\begin{center}
\noindent
last updated Feb 5, 2006
\end{center}
\begin{abstract}
This talk is an introduction to ideas of non-commutative geometry and star products. 
We will discuss consequences for physics in two different settings: 
quantum field theories and astrophysics. In case of quantum field theory, we will discuss two 
recently introduced models in some detail. Astrophysical aspects will be discussed considering
modified dispersion relations.
\end{abstract}

\end{titlepage}

%%%%%%%
%%%
%%% intro
%%%
%%%%%%%

\section{Why Non-Commutativity?}

Non-commutative spaces have a long history. Even in the early days of quantum
mechanics and quantum field theory, continuous space-time and Lorentz symmetry
was considered inappropriate to describe the small scale structure of the
universe \cite{Schroedinger:1934aa}. It was also argued that one should
introduce a fundamental length scale limiting the precision of position
measurements. In \cite{Mach:1937aa,Heisenberg:1938aa} the introduction of a
fundamental length is suggested to cure the ultraviolet divergencies occuring in
quantum field theory. H.~Snyder was the first to formulate these ideas
mathematically \cite{Snyder:1947qz}. He introduced non-commutative coordinates.
Therefore a position uncertainty arises naturally. The success of the
renormalisation programme made people forget about these ideas for some time. But
when the quantisation of gravity was considered thoroughly, it became clear that
the usual concepts of space-time are inadequate and that space-time has to be
quantised or non-commutative, in some way.

There is a deep conceptual difference between quantum field theory and gravity:
In the former space and time are considered as parameters, in the latter as dynamical 
entities.
In order to combine quantum theory and gravitation (geometry), one has to
describe both in the same language, this is the language of algebras
\cite{Majid:1999td}. Geometry can be formulated algebraically in terms of
abelian $\mathbf{C}^*$ algebras and can be generalised to non-abelian $\mathbf{C}^*$
algebras (non-commutative geometry). Quantised gravity 
may even act as a regulator of quantum field theories. This is encouraged by the
fact that non-commutative geometry introduces a lower limit for the precision of
position measurements. There is also a very nice argument showing that, on a
classical level, the selfenergy of a point particle is regularised by the
presence of gravity \cite{Ashtekar:1991hf}. Let us consider an electron and a
shell of radius $\epsilon$ around the electron. The selfenergy of the electron
is the selfenergy of the shell $m(\epsilon)$, in the limit $\epsilon\to 0$.
$m(\epsilon)$ is given by
$$
m(\epsilon) = m_0 + {e^2 \over \epsilon},
$$ 
where $m_0$ is the restmass and $e$ the charge of the electron. In the limit
$\epsilon\to 0$, $m(\epsilon)$ will diverge. Including Newtonian gravity we have
to modify this equation,
$$
m(\epsilon) = m_0 + {e^2 \over \epsilon} - { G m_0^2 \over \epsilon },
$$
$G$ denotes Newton's gravitational constant.
$m(\epsilon)$ will still diverge for $\epsilon\to 0$, unless the mass and the
charge are fine tuned. Considering general relativity, we know that energy,
therefore also the energy of the electron's electric field, is
the source of a gravitational field. Again, we have to modify the above equation,
$$
\label{I.3}
m(\epsilon) = m_0 + {e^2 \over \epsilon} - { G m(\epsilon)^2 \over \epsilon }.
$$
The solution of this quadratic equation is straight forward,
$$
m(\epsilon) = - { \epsilon \over 2G } \pm { \epsilon \over 2G } \sqrt{ 1 + { 4G
\over \epsilon } ( m_0 + { e^2 \over \epsilon } ) }.
$$
We are interested in the positive root. Miraculously, the limit $\epsilon\to 0$
is finite,
$$
\label{I.4}
m(\epsilon\to 0) = { e \over \sqrt{G} }.
$$
This is a non-perturbative result, since $m(\epsilon\to 0)$ cannot be expanded
around $G=0$. $m(\epsilon\to 0)$ does not depend on $m_0$, therefore there is no
fine tuning present. Classical gravity regularises the selfenergy of the electron, on a
classical level. However, this does not make the quantisation of space-time
unnecessary, since quantum corrections to the above picture will again introduce
divergencies. But it provides an example for the regularisation of physical
quantities by introducing gravity. So hope is raised that the introduction of
gravity formulated in terms of non-commutative geometry will regularise physical
quantities even on the quantum level.

On the other hand, 
there is an old and simple argument that a smooth space-time manifold contradicts 
quantum physics. If one localises an event within a
region of extension $l$, an energy of the order $hc/l$ is transfered. This energy 
generates a gravitational field. A strong gravity field
prevents on the other hand signals to reach an observer. Inserting the energy density 
into Einstein's equations gives a corresponding
Schwarzschild radius $r(l)$. This provides a limit on the smallest possible $l$, 
since it is certainly operational impossible to localise
an event beyond this resulting Planck length. To the best of our knowledge, the first 
time this argument was cast into precise mathematics
was in the work by Doplicher, Fredenhagen and Roberts \cite{Doplicher:1994tu}. They 
obtained what is now called the canonical deformation but averaged over
2-spheres. At which energies this transition to discrete structures might take place, 
or at which energies
non-commutative effects occur is a point much debated on.

From various theories
generalised to non-commutative coordinates, limits on the non-commutative scale
have been derived. These generalisations have mainly considered the socalled
canonical non-commutativity,
$$
\left[ \hat x^i, \hat x^j \right] = i \theta^{ij},
$$
$\theta^{ij}=-\theta^{ji}\in\mathbb{C}$. Let us name a
few estimates of the non-commutativity scale.
A very weak limit on the non-commutative scale $\Lambda_{NC}$ is obtained from 
an additional energy loss in stars due to a coupling of the neutral neutrinos 
to the photon, $\gamma\to \nu \bar \nu$ \cite{Schupp:2002up}. They get
$$
\Lambda_{NC} > 81\, GeV.
$$
The estimate is based on the argument that any new energy loss mechanism must
not exceed the standard neutrino losses from the Standard Model by much. A
similar limit is obtained in \cite{Chaichian:2000si} from the calculation of 
the energy levels of the hydrogen atom and the Lamb shift within 
non-commutative quantum electrodynamics ,
$$
\Lambda_{NC} \gtrsim 10^4\, GeV.
$$
If $\Lambda_{NC} = \mathcal O(TeV)$, measurable effects may occur for the
anomalous magnetic moment of the muon which may account for the
reported discrepancy between the Standard Model prediction and the measured
value \cite{Kersting:2001zz}. Also in cosmology and astrophysics non-commutative
effects might be observable. One suggestion is that the modification of the
dispersion relation due to ($\kappa-$)non-commutativity may explain the time delay of
high energy $\gamma$ rays, e.g., from the active galaxy Makarian 142 
\cite{Amelino-Camelia:1999pm,Amelino-Camelia:1998qp}. We will discuss this point in 
Section~\ref{astro}.
A brief introduction to non-commutative geometry will be provided in
Section~\ref{ncg}. Whereas in Section~\ref{ncqft}, we will concern ourselves with quantum
field theory on non-commutative spaces. We will put emphasis on scalar field theories and 
will only briefly discuss the case of gauge theories.

%%%%
%%
%% non-commutative geometry basics
%%
%%%%

\section{Some Basic Notions of Non-Commutative Geometry}
\label{ncg}

At the moment, there are three major approaches tackling the problem of quantising
gravity: String Theory, Quantum Loop Gravity and Non-Commutative Geometry. Before we
discuss some basic concepts of non-commutative geometry, let us state some advantages and
disadvantages of the other theories, cf. \cite{Nicolai:2005mc}. Background independence will be
a major issue. General Relativity can be described in a coordinate free way. In some cases, theories for gravity are expanded around the Minkowski metric. They explicitly
depend on the background Minkowski metric, i.e., background independence is violated.

In String Theory, the basic constituents are 1-dimensional objects, strings. The
interaction between strings can be symbolised by two dimensional Riemann manifolds with
boundary, e.g. a vertex:
\begin{figure}[h]
	\begin{center}
	\epsfig{file=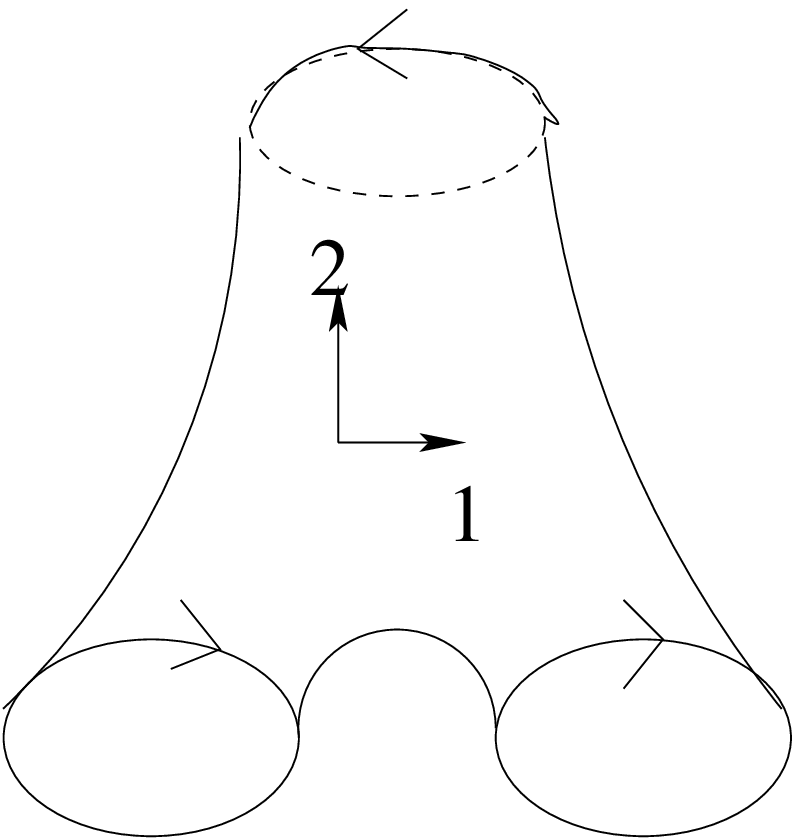, height=3cm}
	\end{center}
	%\caption{}
\end{figure}

\noindent
The interaction region is not a point anymore. Hence, there is also the hope that the
divergencies in the perturbation theory of quantum field theory are not present.\\
\vspace{.7cm}

\begin{tabular}[h]{ll}
Advantages & Disadvantages\\
\\
\hline
\\
graviton contained in the &
higher dimensions needed: \\
particle spectrum & superstrings D=10,11\\
& bosonic string D=26\\
\\
black hole entropy &
dependence on background \\
&space-time geometry\\
\\
mathematical beauty &
many free parameters \\
(dualities, ...) &and string-vacua \\
\\
& almost no predictions

\end{tabular}
\vspace{.7cm}

Quantum Loop Theory studies the canonical quantisation of General Relativity in 3+1
dimensions. \\
\vspace{.7cm}

\begin{tabular}[h]{ll}
Advantages & Disadvantages\\
\\
\hline
\\
background independence &
very few predictions \\
\\
quantised area operator &
no matter included\\
\\
$3+1$ dimensional\\
space-time&
technical difficulties

\end{tabular}
\vspace{.7cm}

We are going to discuss the third approach in more detail in the next subsection. The
three approaches are connected to each other. In \cite{Amelino-Camelia:2003xp}, 
the connection between $\kappa$-deformation
and quantum loop gravity is studied. The authors conclude that the low energy limit of 
quantum loop gravity is a $\kappa$-invariant field theory. This is a far reaching result 
which deserves a lot of attention. Also String theory is related to 
certain non-commutative field theories in the limit of vanishing string coupling 
\cite{Seiberg:1999vs}. 
A better understanding of the interrelations
will provide clues how a proper theory of quantum gravity should look like.

\subsection{Non-Commutative Geometry}

In our approach we consider non-commutative geometry as a generalisation of quantum mechanics.
Thereby, we generalise the canonical commutation relations of the phase space operators $\hat x^i$ and $\hat p_j$.
Most commonly, the commutation relations are 
chosen to be either constant or linear or quadratic in the generators.
In the canonical case the relations are constant,
\be\label{canonical}
[\hat x^i,\hat x^j] = i \theta^{ij},
\ee
where $\theta^{ij}\in\mathbb{C}$ is an antisymmetric matrix, 
$\theta^{ij}=-\theta^{ji}$. The linear or Lie algebra case
\be\label{linear}
[\hat x^i,\hat x^j] = i\lambda^{ij}_k\hat x^k,
\ee
where $\lambda^{ij}_k\in\mathbb{C}$ are the structure constants, basically has
been discussed in two different approaches, fuzzy spheres \cite{Madore:1992bw} and
$\kappa$-deformation \cite{Lukierski:1991pn, Majid:1994cy, Dimitrijevic:2003wv}. Last but not 
least, we have
quadratic commutation relations
\be\label{quadratic}
[\hat x^i,\hat x^j] = ({1\over q}\widehat R^{ij}_{kl}-\delta^i_l
\delta^j_k) \hat x^k \hat x^l,
\ee
where $\widehat R^{ij}_{kl}\in\mathbb{C}$ is the so-called $\widehat R$-matrix. 
For a reference, see e.g., \cite{reshetikhin, Lorek:1997eh}. The relations between coordinates and momenta
(derivatives) can be constructed from the above relations in a consistent way
\cite{Dimitrijevic:2003wv,Dimitrijevic:2004vv}.
Most importantly, usual commutative coordinates are recovered in a certain limit,
$\theta^{ij}\to 0$, $\lambda^{ij}_k\to 0$ or $R^{ij}_{kl}\to 0$, respectively.
In quantum mechanics, the commutation relations lead to the Heisenberg uncertainty,
$$
\Delta x^i \, \Delta p_j \gtrsim \delta^i_j \frac{\hbar}2.
$$
Similarily, we obtain in the non-commutative case a uncertainty relation for the coordinates, e.g.
\be
\Delta x^i \, \Delta x^j \gtrsim \frac{|\theta^{ij}|}2.
\ee

In a next step, we need to know what functions of the non-commutative coordinates are. Classically,
smooth functions can be approximated by power series. So a function $f(x)$ can be written as
$$
f(x) = \sum_I a_I \, (x^1)^{i_1}(x^2)^{i_2}(x^3)^{i_3}(x^4)^{i_4},
$$
where $I=(i_1,..,i_4)$ is a multi-index and considering a four dimensional space. The commutative algebra of functions 
generated by the coordinates $x^1$,
$x^2$, $x^3$ and $x^4$ is denoted by
\be
\mathcal{A} = {\mathbb{C} \langle\langle x^1,...,x^4 \rangle\rangle \over 
[x^i,x^j]} \equiv \mathbb{C} [[x^1,...,x^4]],
\ee
i.e., $[x^i,x^j]=0$. 
The generalisation to the non-commutative algebra of functions $\widehat{\mathcal A}$ on a
non-commutative space
\be
\hat \mathcal A = {\mathbb{C} \langle\langle \hat x^1,...,\hat x^n
\rangle\rangle \over \mathcal I},
\ee
where $\mathcal{I}$ is the ideal generated by the commutation relations of the
coordinate functions, see (\ref{canonical}-\ref{quadratic}). 
Again, an element $\hat f$ of $\hat \mathcal A$ is defined by a power series in the non-commutative 
coordinates.
There is one complication: Since the coordinates do not commute e.g. the monomials $\hat x^i \hat x^j$ 
and $\hat x^j \hat x^i$ are 
different operators. Therefore, we have to specify basis monomials with some care.
That means that we have to give an ordering prescription. Let us discuss two different orderings briefly 
which will be denoted by $:\, :$. Normal ordering means the following:
\bea
\label{normal-ordering}
\nonumber
:\hat x^i: & = & \hat x^i,\, i=1,2,3,4\\ 
:\hat x^2 \hat x^4 \hat x^2 \hat x^1: & = & \hat x^1 (\hat x^2)^2 \hat x^4.
\eea
Powers of $\hat x^1$ come first then powers of $\hat x^2$ and so on. A non-commutative function is given by the formal expansion
\bea
\nonumber
\hat f(\hat x) & = & \sum_I b_I \, :(\hat x^1)^{i_1}(\hat x^2)^{i_2}(\hat x^3)^{i_3}(\hat x^4)^{i_4}:\\
& = & \sum_I b_I \, (\hat x^1)^{i_1}(\hat x^2)^{i_2}(\hat x^3)^{i_3}(\hat x^4)^{i_4}.
\eea
A second choice is the symmetrical ordering. There we define
\bea
\label{symmetrical-ordering}
\nonumber
:\hat x^i: & = & \hat x^i,\\ 
\nonumber
:\hat x^i \hat x^j : & = & \frac12 ( \hat x^i \hat x^j + \hat x^j \hat x^i ),\\
:\hat x^i \hat x^j \hat x^k: & = & \frac16 \bigg( \hat x^i \hat x^j \hat x^k + \hat x^i \hat x^k \hat x^j
	+ \hat x^j \hat x^i \hat x^k\\
\nonumber
&& + \hat x^j \hat x^k \hat x^i + \hat x^k \hat x^i \hat x^j + \hat x^k \hat x^j \hat x^i \bigg),\\
\nonumber
& \vdots &
\eea
 A non-commutative function is given by the formal expansion
\be
\nonumber
\hat f(\hat x) = \sum_I c_I \, :(\hat x^1)^{i_1}(\hat x^2)^{i_2}(\hat x^3)^{i_3}(\hat x^4)^{i_4}:\, .
\ee
Symmetrical ordering can also be achieved by exponentials,
\bea
\nonumber
e^{ik_i \hat x^i} & = & 1+ik_i\hat x^i - \frac12 k_i\hat x^i k_j\hat x^j + \dots = \\
& = & 1+ik_i\hat x^i - \frac12 ( k_1\hat x^1 + \dots + k_4 \hat x^4) ( k_1\hat x^1 + \dots + k_4 \hat x^4) + \dots
\eea
and therefore
\be
\hat f(\hat x) = \int d^4k \, c(k) e^{ik_i\hat x^i},
\ee
with coefficient function $c(k)$. This formula will be of vital importance in the next subsection.

Normal and symmetrical ordering define a different choice of basis in the same non-commutative algebra $\hat \mathcal A$.
Most importantly, many concepts of differential geometry can be formulated using the non-commutative function algebra
$\hat \mathcal A$ such as differential structures.

In the following sections, we will concentrate on the first two cases of non-commutative coordinates, 
namely canonical (\ref{canonical}) and $\kappa$-deformed (\ref{linear})
space-time structures.

\subsection{Star Product}

Star products are a way to return to the familiar concept of commutative functions $f(x)$ within the 
non-commutative realm. In addition, we have to include a non-commutative product denoted by $*$. Earlier we have introduced the
algebras $\mathcal A$ and $\hat \mathcal A$ and have discussed the choice of basis or ordering in the latter. 
We need to establish an isomorphism between the non-commutative algebra $\hat \mathcal A$ and the commutative function algebra
$\mathcal A$.

Let us choose symmetrically ordered monomials as basis in $\hat \mathcal A$. 
Now we map the basis monomials in
$\mathcal{A}$ onto the according symmetrically ordered basis elements of 
$\hat \mathcal A$ 
\bea\label{quantisation}
W : \mathcal{A} & \to & \hat \mathcal{A},\nonumber\\
x^i & \mapsto & \hat x^i,\\
x^i x^j & \mapsto & {1\over 2}( \hat x^i \hat  x^j + \hat x^j \hat x^i)\equiv
	\,\, :\hat x^i \hat x^j:.\nonumber
\eea
The ordering is indicated by $:\,:$.
$W$ is an isomorphism of vector spaces. 
In order to extend {\small $W$} to an algebra isomorphism, we have to 
introduce a new non-commutative multiplication $*$ in $\mathcal{A}$. This 
$*$-product is defined by
\be
\label{star}
W(f*g) := W(f) \cdot W(g) = \hat f\cdot \hat g,
\ee
where $f, g \in \mathcal{A}$, $\hat f, \hat g \in \hat \mathcal{A}$. 
\newpage
Explicitly, we have
\beb
f(x) & = & \sum_I a_I (x^1)^{i_1}(x^2)^{i_2}(x^3)^{i_3}(x^4)^{i_4}\\
& \Vert & \\
& \Vert & \textrm{  Quantisation map W}\\
& \Downarrow & \\
\hat f(\hat x) & = & \sum_I a_I :(\hat x^1)^{i_1}(\hat x^2)^{i_2}(\hat x^3)^{i_3}(\hat x^4)^{i_4}:\, .
\eeb
The star product is constructed in the following way:
\bea
\nonumber
\hat f\cdot \hat g & = & \sum_{I,J} a_I b_J :(\hat x^1)^{i_1}(\hat x^2)^{i_2}(\hat x^3)^{i_3}(\hat x^4)^{i_4}:\, 
:(\hat x^1)^{j_1}(\hat x^2)^{j_2}(\hat x^3)^{j_3}(\hat x^4)^{j_4}:\\
& = & \sum_K c_K :(\hat x^1)^{k_1}(\hat x^2)^{k_2}(\hat x^3)^{k_3}(\hat x^4)^{k_4}:\, ,
\eea
where $\hat g = \sum_J b_J :(\hat x^1)^{j_1}(\hat x^2)^{j_2}(\hat x^3)^{j_3}(\hat x^4)^{j_4}:$. Consequently, we obtain
\be
f * g \, (x) = \sum_J b_J (\hat x^1)^{j_1}(\hat x^2)^{j_2}(\hat x^3)^{j_3}(\hat x^4)^{j_4}\, .
\ee
The information on the non-commutativity of $\hat \mathcal{A}$ is encoded
in the $*$-product.
The Weyl quantisation procedure uses the exponential representation of the symmetrically ordered basis.
The above procedure yields
\bea
\label{invfour}
\hat f=W(f) & = & {1\over (2\pi)^{n/2}} \int d^n\!k\, e^{ik_j \hat x^j} 
	\tilde{f}(k),\\
\label{four}
\tilde f(k) & = & {1\over (2\pi)^{n/2}} \int d^n\!x\, e^{-ik_j x^j} f(x),
\eea
where we have replaced the commutative coordinates by non-commutative ones
($\hat x^i$) 
in the inverse Fourier transformation (\ref{invfour}). 
Hence we obtain
\be
(\mathcal{A},*) \cong (\hat \mathcal{A},\cdot),
\ee
i.e., {\small $W$} is an algebra isomorphism.
Using eqn. (\ref{star}), we are able to compute the star product explicitly,
\be
\label{star2}
W(f*g)={1\over(2\pi)^n} \int d^n\!k\,d^n\!p\,
e^{ik_i\hat x^i}e^{ip_j \hat x^j} \tilde f (k) \tilde g (p).
\ee
Because of the non-commutativity of the coordinates $\hat x^i$, 
we need the Campbell-Baker-Hausdorff (CBH) formula
\be
e^A e^B = e^{A+B +{1\over 2}[A,B] + {1\over 12}[[A,B],B] - 
{1\over 12} [[A,B],A] + \dots}.
\ee
Clearly, we need  to specify the commutation relations of the coordinates in order to evaluate the CBH
formula. We will consider the canonical and the linear case as examples.

\subsubsection{Canonical Case}

Due to the constant commutation relations
$$
{}[\hat x^i, \hat x^j] = i \theta^{ij},
$$
the CBH formula
will terminate, terms with more than one commutator will vanish,
\be
\exp(ik_i \hat x^i) \exp(i p_j \hat x^j) = \exp \left( i(k_i + p_i)\hat x^i - 
{i\over 2}k_i\theta^{ij}p_j \right).
\ee
Eqn. (\ref{star2}) now reads
\be
f*g\,(x) = {1\over (2\pi)^n}\int d^n\!k d^n\!p \,e^{i(k_i+p_i)x^i-{i\over 2}k_i\theta^{ij}p_j}
\tilde f(k)\tilde g(p)
\ee
and we get for the $*$-product the Moyal-Weyl product \cite{Moyal:1949sk}
\be\label{gl3.1}
f*g\, (x) = \exp(\, {i\over 2} {\partial\over \partial x^i}\, \theta^{ij}
\,{\partial\over\partial y^j})\, f(x) g(y)\Big|_{y\to x}\, .
\ee
The same reasoning can be applied to the case of normal ordering. In this basis a non-commutative function $f$
is given by
\be
\hat f(\hat x)= {1\over (2\pi)^{n/2}} \int d^n\!k\, \tilde f(k) e^{ik_1\hat x^1} e^{ik_2\hat x^2} e^{ik_3\hat x^3} e^{ik_4\hat x^4}.
\ee
Equation~(\ref{star2})
has to be replaced by
\be
\hat f\cdot \hat g = {1\over (2\pi)^n}\int d^n\!k d^n\!p\tilde f(k)\tilde g(p) e^{ik_1\hat x^1}\dots e^{ik_4\hat x^4}
e^{ip_1\hat x^1}\dots e^{ip_4\hat x^4}.
\ee
Using CBH formula, $e^{ia\hat x^i}e^{ib\hat x^j}= e^{ib\hat x^j}e^{ia\hat x^i}e^{-iab\theta^{ij}}$, we obtain as a result
for the star product for normal ordering
\be
f*_N g\, (x) = \exp (\sum_{i>j} i {\partial\over \partial x^i}\,\theta^{ij}
\,{\partial\over\partial y^j})\, f(x) g(y)\Big|_{y\to x}\, .
\ee
In both cases, we can now explicitly show that Eq.~(\ref{canonical}) is satisfied.
The star product enjoys a very important property,
\be
\int d^4 x\, f\star g \star h = \int d^4 x \, h\star f\star g,\qquad \int d^4x f\star g = \int d^4 x\, f\cdot g.
\ee
This is called the trace property.

\subsubsection{$\kappa$-Deformed Case}

The following choice of linear commutation relation is called $\kappa$-defomation:
\be
\label{kappa}
[ \hat x^1, \hat x^p] = i a \hat x^p, \qquad
[\hat x^q, \hat x^p] =0,
\ee
where $p,q=2,3,4$. Because the structure is more involved the computation of the star product is not
as easy as in the canonical case. Therefore we will just state the result.
The symmetrically ordered star product is given by \cite{Dimitrijevic:2004vv}
\bea
\label{kappa-star}
f*g\, (x) = \int d^4k \, d^4 p \, \tilde f(k) \tilde g(p)
 \, e^{i(\omega_k + \omega_p)x^1} e^{i\vec x( \vec k e^{a\omega_p} A(\omega_k,\omega_p)+ \vec p A(\omega_p,\omega_k))},
\eea
where $k=(\omega_k,\vec k)$, and $\vec x=(x^2,x^3,x^4)$.
We have used the definition
\be
\label{A}
A(\omega_k,\omega_p) \equiv  \frac{a(\omega_k+\omega_p)}{e^{a(\omega_k+\omega_p)}-1} 
	\frac{e^{a\omega_k}-1}{a\omega_k}.
\ee
The normal ordered star product has the form 
\cite{Dimitrijevic:2004vv}
\bea
f*_N g\, (x) & = & \lim_{\mbox{\footnotesize $\begin{array}{c} y\to x \\ z\to x \end{array}$}}
e^{x^j \frac{\pat}{\pat y^j} ( e^{-ia \frac{\pat}{\pat z^1}} - 1 ) } f(y) g(z)
\nonumber\\
& = & \int \frac{d^4p\, d^4k}{(2\pi)^4} e^{i x^1(\omega_k+\omega_p)}
e^{i\vec x ( \vec k e^{a\omega_p}+\vec p ) } 
\tilde f(k) \tilde g(p).
\eea
In the $\kappa$-deformed case, the trace property is modified. We have to introduce a integration measure $\mu(x)$:
\be
\int d^4 x \mu(x) \, (f\star g)\,(x) = \int d^4 x \mu(x) \, (g\star f)\,(x). 
\ee
The above Equation also determines the function $\mu(x)$, see e.g. \cite{Dimitrijevic:2003wv}.

%%%%
%%
%% ncqft
%%
%%%%

\section{Non-Commutative Quantum Field Theory}
\label{ncqft}

Many models of non-commutative quantum field theory have been studied in recent years,
and a coherent picture is beginning to emerge. One of the surprising features is the socalled
ultraviolet (UV) / infrared (IR) mixing, where the usual divergences of field theory in the UV
are reflected by new singularities in the IR. This is essentially a reflection of the uncertainty
relation: determining some coordinates to very high precision (UV) implies a large uncertainty
(IR) for others. This leads to a serious problem for the usual renormalisation procedure of quantum
field theories, which has only recently been overcome for a scalar field theoretical
model on the canonical deformed Euclidean space \cite{Grosse:2003nw,Grosse:2004yu}. This model will 
be discussed in the Section~\ref{scalar}. Most models constructed so far use canonical space-time, the 
simplest deformation. Therefore we will also describe a quantum field theory on a more complicated structure, namely
a $\kappa$-deformed space, here. Nevertheless the problem of UV/IR mixing could not be solved by
this deformation.

\subsection{Scalar Field Theory}
\label{scalar}

In this subsection, we want to sketch two different models of scalar fields on non-commutative space-time.
The first model is formulated on 
$\kappa$-deformed Euclidean space \cite{Grosse:2005iz} , the second model
given in \cite{Grosse:2003nw,Grosse:2004yu} on canonically deformed Euclidean space.

The commutation relations of the coordinates for the $\kappa$-deformed case are given by Eq.~(\ref{kappa}).
For simplicity, we concentrate on the Euclidean version and use the 
symmetrically ordered star product is given in Eq.~(\ref{kappa-star}). $\kappa$-deformed spaces allow for 
a generalised coordinate symmetry, socalled $\kappa$-Poincar\'e symmetry \cite{Lukierski:1991pn,
Dimitrijevic:2003wv}. Therefore, also the action 
should be invariant under this symmetry. In \cite{Dimitrijevic:2003wv}, the $\kappa$-Poincar\'e algebra and
the action of its generators on commutative functions is explicitly calculated starting at the commutation
relations (\ref{kappa}). In order to describe scalar fields on a $\kappa$-deformed space we need to write down
an action. Therefore, we have to know the $\kappa$-deformed version of the Klein-Gordon operator and an
integral invariant under $\kappa$-Poincar\'e transformations.
%Let us state a very useful identity which we will need a lot in  the calculations:
%\be
%\label{id1}
%e^{-a\omega_2} A(-\omega_1,-\omega_2) = A(\omega_1,\omega_2).
%\ee
The Klein-Gordon operator is a Casimir in the translation generators (momenta) \cite{Dimitrijevic:2003wv}. 
Acting on commutative functions, it is given by the following expression:
\be
\Box^* = \sum_{i=1}^4 \pat_i\pat_i \frac{2(1-\cos a\pat_1)}
	{a^2\pat_1^2}.
\ee
A $\kappa$-Poincar\'e invariant integral is provided in \cite{Moller:2004sk} and has the following form:
\be
(\phi, \psi) = \int d^4 x \phi ( K \bar\psi),
\ee
where $K$ is a suitable differential operator,
\be
K = \left( \frac{ -ia\pat_1}{e^{-ia\pat_1}-1} \right)^{3}.
\ee
In momentum space, this amounts to 
\be
(\phi,\psi) = \int d^4 q \left( \frac{-a\omega_q}{e^{-a\omega_q}-1}\right)^3 \, \tilde \phi(q) \bar{\tilde\psi}(q).
\ee
And therefore, the action for a scalar field with $\phi^4$ interaction is given by
\bea
\label{dint}
S[\phi] & = & -(\phi, (\Box^* -m^2)\psi) \\
\nonumber
&& +\,  \frac{g}{4!} \left( b(\, \phi*\phi, \, \phi*\phi)
%+  c (\phi*\phi*\phi,\,  \phi)  
+ d (\,\phi*\phi*\phi*\phi,1)
\right).
\eea
In the above action, we have not included all possible interaction terms. A term proportional to 
$(\phi\star\phi\star\phi,\, \phi)$ is missing. This term, however, would lead to a peculiar behaviour, therefore
it is ignored. For more details see \cite{Grosse:2005iz}.

In momentum space, the action has the following form:
\bea
\label{action-kappa}
S[\phi] & = &  \int d^4q \left( \frac{-a\omega_q}{e^{-a\omega_q}-1}\right)^3
\tilde \phi(q) \left(
q^2 \frac{2(\cosh a\omega_q - 1)}{a^2 \omega_q^2} + m^2
\right)
\bar{\tilde\phi}(q)
\\
\nonumber
&& 
\hspace{-.2cm}
+ b\,\frac{g}{4!} \int d^4 z  \prod_{i=1}^4 d^4 k_i
\left( \frac{a(\omega_{k_3}+\omega_{k_4})}{e^{a(\omega_{k_3}+\omega_{k_4})}-1} \right)^3
\, 
\tilde\phi(k_1)\tilde\phi(k_2) \tilde\phi(k_3)\tilde\phi(k_4) \\
\nonumber
&& \times
	e^{i z^1 \sum \omega_{k_i}}\,
	exp\bigg(i\vec z \left[
	\vec k_1 e^{a\omega_{k_2}} A(\omega_{k_1},\omega_{k_2}) + 
	\vec k_2 A(\omega_{k_2},\omega_{k_1})
\right.
\\ 
\nonumber && \hspace{.5cm} \left.+
	\vec k_3 e^{-a\omega_{k_4}} A(-\omega_{k_3},-\omega_{k_4}) +
	\vec k_4 A(-\omega_{k_4},-\omega_{k_3})
	\right] \bigg)
\nonumber
\\
\nonumber
&& 
\hspace{-.2cm}
+ d\,\frac{g}{4!} \int d^4 z  \prod_{i=1}^4 d^4 k_i
\, e^{i z^1 \sum \omega_{k_i}}
\tilde\phi(k_1)\tilde\phi(k_2) \tilde\phi(k_3)\tilde\phi(k_4) \\
\nonumber
&& \times
	e^{i\vec z 
	( \vec k_1 e^{a\omega_{k_2}}  A(\omega_{k_1},\omega_{k_2}) 
		+ \vec k_2 A( \omega_{k_2}, \omega_{k_1}) ) e^{a(\omega_{k_3}+\omega_{k_4})} 
		A(\omega_{k_1}+\omega_{k_2},\omega_{k_3}+\omega_{k_4}) }
		\\
\nonumber
&& \times
	e^{i \vec z ( \vec k_3 e^{a\omega_{k_4}}  A(\omega_{k_3},\omega_{k_4}) 
		+ \vec k_4 A( \omega_{k_4}, \omega_{k_3}) ) A(\omega_{k_3}+\omega_{k_4},\omega_{k_1}+\omega_{k_2})
	}.
\eea
Note that $ \bar{\tilde \phi}(k) = \tilde \phi(-k)$, for real fields $\phi(x)$. The $x$-dependent
phase factors are a direct result of the star product (\ref{kappa-star}), $b$ and $d$ are real parameters. In the 
case of canonical deformation, the phase factor is independent of $x$. 
Similar to the commutative case, we want to extract amplitudes for Feynman diagrams from a generating
functional by differentiation.
The generating functional can be defined as 
\be
\label{f1}
Z_\kappa[J] = \int \mathcal D\phi e^{-S[\phi] + \frac12 (J,\phi) + \frac12 (\phi,J) }.
\ee
The $n$-point functions $\tilde G_n(p_1,\dots,p_n)$ are given by functional
differentiation:
\be
\label{f2}
\tilde G_n(p_1,\dots,p_n) = \frac{\delta^n}{\delta \tilde J(-p_1) \dots \delta
\tilde J (-p_n) } Z_\kappa [J]\Big|_{J=0}.
\ee
Let us first consider the free case. For the free generating functional
$Z_{0,\kappa}$ we obtain from Eq.~(\ref{f1})
\bea
\nonumber
Z_{0,\kappa}[J] & = & \int \mathcal D\phi \exp \left[
-\frac12 \int d^4k \left( \frac{-a \omega_k}{e^{-a\omega_k}-1} \right)^3
\tilde\phi (k) ( \mathcal M_k + m^2 ) \tilde \phi(-k)
	\right.\\
\label{f3}&& \hspace{-.5cm}
\left.  +\frac12 \int d^4k \left( \left( \frac{-a \omega_k}{e^{-a\omega_k}-1}
\right)^3+ \left( \frac{a \omega_k}{e^{a\omega_k}-1} \right)^3 \right) \tilde
J(k) \, \tilde \phi(-k)
	\right],
\eea
where we have defined
\be
\mathcal M_k := \frac{ 2 k^2 (\cosh a\omega_k -1)}{a^2 \omega_k^2}.
\ee
The same manipulations as in the classical case yield
\be
\label{f4}
Z_{0,\kappa}[J] =Z_{0,\kappa}[0] e^{ \frac12 \int d^4k 
\left( \frac{-a \omega_k}{e^{-a\omega_k}-1} \right)^3 \frac{\tilde J(k) 
\tilde J(-k)}{\mathcal M_k+m^2}}.
\ee
We will always consider the normalised functional, which we obtain by dividing
with $Z_{0,\kappa}[0]$.
Now, the free propagator is given by
\bea
\label{f5}
\tilde G(k,p) & = & \frac{\delta^2}{\delta \tilde J(-k) \delta \tilde J(-p)}
Z_{0,\kappa}[J]\Big|_{J=0}\\
\nonumber & = & 
L(\omega_k) \frac{\delta^{(4)} (k+p)}{\mathcal M_k + m^2}\equiv \delta^{(4)}(k+p)Q_k. 
\eea
For the sake of brevity, we have introduced
\be
\label{f6}
L(\omega_k) := \frac12 \left( 
\left( \frac{-a \omega_k}{e^{-a\omega_k}-1} \right)^3 +
\left( \frac{a \omega_k}{e^{a\omega_k}-1} \right)^3 \right).
\ee
We can rewrite the full generating functional in the form
\be
\label{f7}
Z_\kappa[J] = e^{-S_I[ 1/L(\omega_k) \frac{\delta}{\delta \tilde J(-k)}]}
Z_{0,\kappa}[J].
\ee
The full propagator to first order in the coupling parameter is given by the
connected part of the expression
\be
\label{f11}
\tilde G^{(2)}(p,q) = \frac{\delta^2}{\delta \tilde J(-p) \delta \tilde J(-q)}
Z_\kappa[J]\Big|_{J=0}. 
\ee
The aim ist to compute tadpole diagram contributions. In order
to do so, we expand the generating functional (\ref{f7}) in powers of the
coupling constant $g$. Using Eq.~(\ref{action}), we obtain
\be
\label{f8}
Z_\kappa[J] = Z_{0,\kappa}[J] + Z_\kappa^1[J] + \mathcal O(g^2).
\ee
The details of the calculation are given in \cite{Grosse:2005iz}. Let us just state the results.
As in the canonically deformed case, we can distinguish between planar and non-planar diagrams.
The planar diagrams display a linear UV divergence. The non-planar diagrams are finite for generic
external momenta, $p$ and $p$, resp. However, in the exceptional case $\omega_p=\omega_k=0$ the 
amplitudes also diverge linearly in the UV cut-off.

Let us switch the second model. Remarkably, the problem of UV/IR divergencies is solved in this case. 
And the model turns out to be renormalisable.
We will briefly sketch the model ant its peculiarities.
Again, it is a scalar field
theory. It is defined on the 4-dimensional quantum plane $\mathbb R^4_\theta$,
with commutation relations $[x_\mu,x_\nu] = i \theta_{\mu\nu}$.
The UV/IR mixing was taken into account  through a
modification of the free Lagrangian, by adding an oscillator term
which modifies the spectrum of the free action:
\be S= \int
d^4x \Big( \frac{1}{2} \partial_\mu \phi \star
\partial^\mu \phi + \frac{\Omega^2}{2} (\tilde{x}_\mu \phi) \star
(\tilde{x}^\mu \phi) + \frac{\mu^2}{2} \phi \star \phi +
\frac{\lambda}{4!} \phi \star \phi \star \phi \star
\phi\Big)(x)\;. \label{action} 
\ee 
Here $\star$ is the Moyal star product (\ref{gl3.1}).
The harmonic oscillator term in Eq.~(\ref{action}) 
was found as a result of the renormalisation proof. The model
is covariant under
the Langmann-Szabo duality relating short distance
and long distance behaviour.  

The renormalizstion proof proceeds by using a matrix base $b_{nm}$. The remarkable feature of this base is 
that the star product is reduced to a matrix product,
\be
b_{kl} \star b_{mn} = \delta_{lm} b_{kn}.
\ee 
We can expand the fields in terms of this base:
\be
\phi=\sum_{m,n} \phi_{nm} b_{nm}(x).
\ee
This leads to a dynamical matrix model of the type:
\be
S  =(2\pi\theta)^2 \sum_{m,n,k,l\in \mathbb{N}^2}
\Big(\frac{1}{2} \phi_{mn} \Delta_{mn;kl} \phi_{kl} +
\frac{\lambda}{4!} \phi_{mn}\phi_{nk} \phi_{kl} \phi_{lm}\Big)\;,
\label{Sm}
\ee
where
\bea
\Delta_{\stackrel{m^1}{m^2}\stackrel{n^1}{n^2};\stackrel{k^1}{k^2}\stackrel{l^1}{l^2}} & = &
\big(\mu^2{+} \frac{2{+}2\Omega^2}{\theta}
(m^1{+}n^1{+}m^2{+}n^2{+}2) \big) \delta_{n^1k^1} \delta_{m^1l^1}
\delta_{n^2k^2} \delta_{m^2l^2} \nonumber
\\
&& \hspace{-2.6cm}
- \frac{2{-}2\Omega^2}{\theta} \big(\sqrt{k^1l^1}\,
  \delta_{n^1+1,k^1}\delta_{m^1+1,l^1} + \sqrt{m^1n^1}\,
  \delta_{n^1-1,k^1} \delta_{m^1-1,l^1}\big)\delta_{n^2k^2}
  \delta_{m^2l^2}
\nonumber
\\
&&\hspace{-2.6cm}
- \frac{2{-}2\Omega^2}{\theta} \big(\sqrt{k^2l^2}\,
  \delta_{n^2+1,k^2}\delta_{m^2+1,l^2} + \sqrt{m^2n^2}\,
  \delta_{n^2-1,k^2} \delta_{m^2-1,l^2}\big)\delta_{n^1k^1}
  \delta_{m^1l^1} .
\label{Gm}
\eea
The interaction part becomes a trace of product of matrices, and no
oscillations occur in this basis. In the $\kappa$-deformed case, we have discussed before,
$x-$dependent phases occurred. 
Here, the interaction terms have a very simple structure, but
the propagator obtained from the free part is quite complicated.
For the details see \cite{Grosse:2004yu}.

These propagators show asymmetric decay
properties: they decay exponentially on particular directions, but
have power law decay in others.  These decay properties
are crucial for the perturbative
renormalizability respectively nonrenormalizability of the models.
The renormalisation proof follows the ideas of Polchinski 
\cite{Polchinski:1983gv}. 
The integration of the
Polchinski equation from some initial scale down to the
renormalisation scale leads to divergences after removing the
cutoff. For relevant/marginal operators, one therefore has to fix 
certain initial conditions. The goal is then to find a procedure involving only
a finite number of such operators. Through the invention of
a mixed integration procedure and by proving a certain power
counting theorem, they were able to reduce the divergences 
to only four relevant/marginal 
operators. A somewhat 
long sequence of estimates and arguments then leads to the proof of
renormalisation. Afterwards, they could also derive $\beta$-functions 
for the coupling constant flow, which shows that the
ratio of the coupling constants $\lambda / \Omega ^2$ remains
bounded along the renormalisation group flow up to first order. 
The renormalisability of this model is a very important result and so far the 
only example of a renormalisable non-commutative model.

\subsection{Gauge Theories}

At present, particles and their interactions are described by gauge theories. 
The most prominent gauge theory is the Standard Model, which incudes the electromagnetic force
and the strong and weak nuclear forces.
Therefore, it is of vital interest to extend the ideas of non-commutative geometry and
the renormalisation 
method described above to gauge field theories.
Let us sketch two approaches:
\begin{enumerate}
\item 
Non-commutative gauge theories can be formulated by introducing socalled Seiberg-Witten maps 
\cite{Seiberg:1999vs,Madore:2000en}. There, the non-commutative gauge fields are given as a power series in 
the non-commutativity parameters. They depend on the commutative gauge field and gauge parameter and are solutions
of gauge equivalence conditions. Therefore, no additional degrees of freedom are introduced. A major advantage of this
approach is that there are no limitations to the gauge group. For an introduction see e.g. \cite{Wohlgenannt:2003de}.
The Standard Model of elementary particle physics is discussed in \cite{Calmet:2001na,Melic:2005am,Melic:2005fm} using
this approach. However, these theories seemingly have to be considered as effective theories, since in 
\cite{Wulkenhaar:2001sq} the non-renormalisability of non-commutative QED has explicitly been shown.

\item
A second approach  starts from covariant coordinates 
$B_\mu = \theta^{-1}_{\mu\nu} x^\nu + A_\mu$ \cite{Madore:2000en}. These objects transform covariantly under
gauge transformations:
$$
B_\mu \to U^* \star B_\mu \star U,
$$
with $U^* \star U = U\star U^* = 1$. This is analogous to the introduction of covariant derivatives. Covariant coordinates only exist on
non-commutative spaces.
We can write down a gauge invariant version of the action~(\ref{action}):
\be
S = \int d^4\,  \left( \frac12 \phi\star [B_\nu\stackrel{\star}{,} [B^\nu \stackrel{\star}{,} \phi]] + 
\frac{\Omega^2}2\phi \star B_\nu\stackrel{\star}{,} \{ \{B^\nu \stackrel{\star}{,} \phi\}\} \right),
\ee
where we have used $[x^\mu\stackrel{\star}{,}f]=i\theta^{\mu\nu}\partial_\nu f$.
\end{enumerate}

\section{Astrophysical Considerations}
\label{astro}

In this Section, we want to discuss the modification of dispersion relations in 
$\kappa$-deformed space-time. This modifications lead to a bound on the non-commutativity parameter.
We will follow the presentation given in \cite{Amelino-Camelia:1999pm}.
Similar considerations can be found e.g. in \cite{Biller:1998hg,Tamaki:2002iz}.

In Section~\ref{ncqft}, we have discussed a scalar model on a $\kappa$-deformed Euclidean space.
Here, we consider $\kappa$-Minkowski space-time with the relations
\be
[\hat x^i, \hat t] = i \lambda \hat x^i, \qquad [\hat x^i,\hat x^j]=0,
\ee
$i,j=1,2,3$. The modification of the dispersion relation is to the modified D'Alembert Operator 
also briefly discussed in Section~\ref{ncqft}:
\be
\lambda^{-2} \left( e^{\lambda \omega} + e^{-\lambda \omega} -2 \right) -{\vec k}^2e^{-\lambda \omega} = m^2.
\ee
In the commutative limit, $\lambda \to 0$, we of course obtain the usual relation
$$
\omega^2 - {\vec k}^2 = m^2\, .
$$
The velocity for a massless particle is given by
\be
\vec v = \frac{d\omega}{d\vec k} = \frac{\lambda \vec k}{\lambda^2 {\vec k}^2 + 
\frac{\lambda\omega}{|\lambda \omega |}\sqrt{\lambda^2{\vec k}^2}}.
\ee
For the speed one obtains
\be
v=e^{-\lambda\omega}\approx 1-\lambda \omega.
\ee
That means that the velocity of the particle depends on its energy. Particles
with different energies will take a different amount of time for the same distance.

Let us consider $\gamma$-ray bursts form active galaxies, such as Makarian 142.
The time difference $\delta t$ in the time of arrival of photons with different energy can therefore be estimated as
\be
|\delta t| \approx \lambda \frac{L}{c} \delta \omega,
\ee
where $L$ is the distance of the galaxy, $\delta \omega$ the energy range of the burst and $ \lambda$ the 
non-commutativity parameter. A usual $\gamma$-ray burst spreads over a range of $0.1\, -\, 100$ MeV.
Data already available seem to imply that $\lambda < 10^{-33} m$.

%%%%
%%
%% acknowledgement
%%
%%%%

\subsection*{Acknowledgement}

This work has been supported by FWF (Austrian Science Fund), project P16779-N02 and by the
Austro-Ukrainian Institute for Science and Technology (AUI).

\bibliographystyle{../latex-styles/utphys}
\bibliography{../tuw}

\providecommand{\href}[2]{#2}\begingroup\raggedright\begin{thebibliography}{10}

\bibitem{Schroedinger:1934aa}
E.~Schr{\"o}dinger, ``{\"U}ber die {U}nanwendbarkeit der {Geometrie im
  Kleinen},'' {\em Die Naturwiss.} {\bf 31} (1934) 518.

\bibitem{Mach:1937aa}
A.~Mach, ``{Die Geometrie kleinster R\"aume. I},'' {\em Z.Phys.} {\bf 104}
  (1937) 93.

\bibitem{Heisenberg:1938aa}
W.~Heisenberg, ``{\"U}ber die in der {Theorie der Elementarteilchen auftretende
  universelle} {L\"a}nge,'' {\em Ann.Phys.} {\bf 32} (1938) 20.

\bibitem{Snyder:1947qz}
H.~S. Snyder, ``{Quantized Space-Time},'' {\em Phys. Rev.} {\bf 71} (1947)
38--41.
%%CITATION = PHRVA,71,38;%%.

\bibitem{Majid:1999td}
S.~Majid, ``Quantum groups and noncommutative geometry,'' {\em J. Math. Phys.}
  {\bf 41} (2000) 3892--3942,
\href{http://www.arXiv.org/abs/hep-th/0006167}{{\tt hep-th/0006167}}.
%%CITATION = HEP-TH 0006167;%%.

\bibitem{Ashtekar:1991hf}
A.~Ashtekar, ``Lectures on nonperturbative canonical gravity,''. Singapore,
  Singapore: World Scientific (1991) 334 p. (Advanced series in astrophysics
  and cosmology, 6), Chapter 1.

\bibitem{Doplicher:1994tu}
S.~Doplicher, K.~Fredenhagen, and J.~E. Roberts, ``The quantum structure of
  space-time at the planck scale and quantum fields,'' {\em Commun. Math.
  Phys.} {\bf 172} (1995) 187--220,
\href{http://www.arXiv.org/abs/hep-th/0303037}{{\tt hep-th/0303037}}.
%%CITATION = HEP-TH 0303037;%%.

\bibitem{Schupp:2002up}
P.~Schupp, J.~Trampeti\v{c}, J.~Wess, and G.~Raffelt, ``The photon neutrino
  interaction in non-commutative gauge field theory and astrophysical bounds,''
  {\em Eur. Phys. J.} {\bf C36} (2004) 405--410,
\href{http://www.arXiv.org/abs/hep-ph/0212292}{{\tt hep-ph/0212292}}.
%%CITATION = HEP-PH 0212292;%%.

\bibitem{Chaichian:2000si}
M.~Chaichian, M.~M. Sheikh-Jabbari, and A.~Tureanu, ``Hydrogen atom spectrum
  and the {Lamb} shift in noncommutative {QED},'' {\em Phys. Rev. Lett.} {\bf
  86} (2001) 2716,
\href{http://www.arXiv.org/abs/hep-th/0010175}{{\tt hep-th/0010175}}.
%%CITATION = HEP-TH 0010175;%%.

\bibitem{Kersting:2001zz}
N.~Kersting, ``$(g-2)\mu$ from noncommutative geometry,'' {\em Phys. Lett.}
  {\bf B527} (2002) 115--118,
\href{http://www.arXiv.org/abs/hep-ph/0109224}{{\tt hep-ph/0109224}}.
%%CITATION = HEP-PH 0109224;%%.

\bibitem{Amelino-Camelia:1999pm}
G.~Amelino-Camelia and S.~Majid, ``Waves on noncommutative spacetime and
  gamma-ray bursts,'' {\em Int. J. Mod. Phys.} {\bf A15} (2000) 4301--4324,
\href{http://www.arXiv.org/abs/hep-th/9907110}{{\tt hep-th/9907110}}.
%%CITATION = HEP-TH 9907110;%%.

\bibitem{Amelino-Camelia:1998qp}
G.~Amelino-Camelia, J.~R. Ellis, N.~E. Mavromatos, D.~V. Nanopoulos, and
  S.~Sarkar, ``Sensitivity of astrophysical observations to gravity- induced
  wave dispersion in vacua,''
\href{http://www.arXiv.org/abs/astro-ph/9810483}{{\tt astro-ph/9810483}}.
%%CITATION = ASTRO-PH 9810483;%%.

\bibitem{Nicolai:2005mc}
H.~Nicolai, K.~Peeters, and M.~Zamaklar, ``Loop quantum gravity: An outside
  view,'' {\em Class. Quant. Grav.} {\bf 22} (2005) R193,
\href{http://www.arXiv.org/abs/hep-th/0501114}{{\tt hep-th/0501114}}.
%%CITATION = HEP-TH 0501114;%%.

\bibitem{Amelino-Camelia:2003xp}
G.~Amelino-Camelia, L.~Smolin, and A.~Starodubtsev, ``Quantum symmetry, the
  cosmological constant and {P}lanck scale phenomenology,'' {\em Class. Quant.
  Grav.} {\bf 21} (2004) 3095--3110,
\href{http://www.arXiv.org/abs/hep-th/0306134}{{\tt hep-th/0306134}}.
%%CITATION = HEP-TH 0306134;%%.

\bibitem{Seiberg:1999vs}
N.~Seiberg and E.~Witten, ``String theory and noncommutative geometry,'' {\em
  JHEP} {\bf 09} (1999) 032,
\href{http://arXiv.org/abs/hep-th/9908142}{{\tt hep-th/9908142}}.
%%CITATION = HEP-TH 9908142;%%.

\bibitem{Madore:1992bw}
J.~Madore, ``The {Fuzzy Sphere},'' {\em Class. Quant. Grav.} {\bf 9} (1992)
69--88.
%%CITATION = CQGRD,9,69;%%.

\bibitem{Lukierski:1991pn}
J.~Lukierski, H.~Ruegg, A.~Nowicki, and V.~N. Tolstoi, ``{q} deformation of
  {Poincar\'{e}} algebra,'' {\em Phys. Lett.} {\bf B264} (1991)
331--338.
%%CITATION = PHLTA,B264,331;%%.

\bibitem{Majid:1994cy}
S.~Majid and H.~Ruegg, ``Bicrossproduct structure of $\kappa$-{Poincar\'{e}}
  group and noncommutative geometry,'' {\em Phys. Lett.} {\bf B334} (1994)
  348--354,
\href{http://arXiv.org/abs/hep-th/9405107}{{\tt hep-th/9405107}}.
%%CITATION = HEP-TH 9405107;%%.

\bibitem{Dimitrijevic:2003wv}
M.~Dimitrijevi\'c, L.~Jonke, L.~M{\"o}ller, E.~Tsouchnika, J.~Wess, and
  M.~Wohlgenannt, ``Deformed {Field Theory} on $\kappa$-spacetime,'' {\em Eur.
  Phys. J.} {\bf C31} (2003) 129--138,
\href{http://www.arXiv.org/abs/hep-th/0307149}{{\tt hep-th/0307149}}.
%%CITATION = HEP-TH 0307149;%%.

\bibitem{reshetikhin}
N.~Reshetikhin, L.~Takhtadzhyan, and L.~Faddeev, ``Quantization of {Lie} groups
  and {Lie} algebras,'' {\em Leningrad Math. J.} {\bf 1} (1990) 193.

\bibitem{Lorek:1997eh}
A.~Lorek, W.~Weich, and J.~Wess, ``Non-commutative {Euclidean} and {Minkowski}
  structures,'' {\em Z. Phys.} {\bf C76} (1997) 375--386,
\href{http://arXiv.org/abs/q-alg/9702025}{{\tt q-alg/9702025}}.
%%CITATION = Q-ALG 9702025;%%.

\bibitem{Dimitrijevic:2004vv}
M.~Dimitrijevi\'c, L.~M{\"o}ller, and E.~Tsouchnika, ``Derivatives, forms and
  vector fields on the $\kappa$-deformed {E}uclidean space,'' {\em J. Phys.}
  {\bf A37} (2004) 9749--9770,
\href{http://www.arXiv.org/abs/hep-th/0404224}{{\tt hep-th/0404224}}.
%%CITATION = HEP-TH 0404224;%%.

\bibitem{Moyal:1949sk}
J.~E. Moyal, ``Quantum mechanics as a statistical theory,'' {\em Proc.
  Cambridge Phil. Soc.} {\bf 45} (1949)
99--124.
%%CITATION = PCPSA,45,99;%%.

\bibitem{Grosse:2003nw}
H.~Grosse and R.~Wulkenhaar, ``Renormalisation of {$\phi^4$} theory on
  noncommutative {$\mathbb R^2$} in the matrix base,'' {\em JHEP} {\bf 12}
  (2003) 019,
\href{http://www.arXiv.org/abs/hep-th/0307017}{{\tt hep-th/0307017}}.
%%CITATION = HEP-TH 0307017;%%.

\bibitem{Grosse:2004yu}
H.~Grosse and R.~Wulkenhaar, ``Renormalisation of {$\phi^4$} theory on
  noncommutative {$\mathbb R^4$} in the matrix base,'' {\em Commun. Math.
  Phys.} {\bf 256} (2005) 305--374,
\href{http://www.arXiv.org/abs/hep-th/0401128}{{\tt hep-th/0401128}}.
%%CITATION = HEP-TH 0401128;%%.

\bibitem{Grosse:2005iz}
H.~Grosse and M.~Wohlgenannt, ``On $\kappa$-deformation and {UV/IR} mixing,''
\href{http://www.arXiv.org/abs/hep-th/0507030}{{\tt hep-th/0507030}}.
%%CITATION = HEP-TH 0507030;%%.

\bibitem{Moller:2004sk}
L.~M{\"o}ller, ``A symmetry invariant integral on $\kappa$-deformed
  spacetime,''
\href{http://www.arXiv.org/abs/hep-th/0409128}{{\tt hep-th/0409128}}.
%%CITATION = HEP-TH 0409128;%%.

\bibitem{Polchinski:1983gv}
J.~Polchinski, ``Renormalization and effective lagrangians,'' {\em Nucl. Phys.}
  {\bf B231} (1984)
269--295.
%%CITATION = NUPHA,B231,269;%%.

\bibitem{Madore:2000en}
J.~Madore, S.~Schraml, P.~Schupp, and J.~Wess, ``Gauge theory on noncommutative
  spaces,'' {\em Eur. Phys. J.} {\bf C16} (2000) 161--167,
\href{http://www.arXiv.org/abs/hep-th/0001203}{{\tt hep-th/0001203}}.
%%CITATION = HEP-TH 0001203;%%.

\bibitem{Wohlgenannt:2003de}
M.~Wohlgenannt, ``Introduction to a non-commutative version of the standard
  model,''
\href{http://www.arXiv.org/abs/hep-th/0302070}{{\tt hep-th/0302070}}.
%%CITATION = HEP-TH 0302070;%%.

\bibitem{Calmet:2001na}
X.~Calmet, B.~Jur\v{c}o, P.~Schupp, J.~Wess, and M.~Wohlgenannt, ``The standard
  model on non-commutative space-time,'' {\em Eur. Phys. J.} {\bf C23} (2002)
  363--376,
\href{http://arXiv.org/abs/hep-ph/0111115}{{\tt hep-ph/0111115}}.
%%CITATION = HEP-PH 0111115;%%.

\bibitem{Melic:2005am}
B.~Melic, K.~Passek-Kumericki, J.~Trampetic, P.~Schupp, and M.~Wohlgenannt,
  ``The standard model on non-commutative space-time: Strong interactions
  included,'' {\em Eur. Phys. J.} {\bf C42} (2005) 499--504,
\href{http://www.arXiv.org/abs/hep-ph/0503064}{{\tt hep-ph/0503064}}.
%%CITATION = HEP-PH 0503064;%%.

\bibitem{Melic:2005fm}
B.~Melic, K.~Passek-Kumericki, J.~Trampetic, P.~Schupp, and M.~Wohlgenannt,
  ``The standard model on non-commutative space-time: Electroweak currents and
  higgs sector,'' {\em Eur. Phys. J.} {\bf C42} (2005) 483--497,
\href{http://www.arXiv.org/abs/hep-ph/0502249}{{\tt hep-ph/0502249}}.
%%CITATION = HEP-PH 0502249;%%.

\bibitem{Wulkenhaar:2001sq}
R.~Wulkenhaar, ``Non-renormalizability of $\theta$-expanded noncommutative
  {QED},'' {\em JHEP} {\bf 03} (2002) 024,
\href{http://www.arXiv.org/abs/hep-th/0112248}{{\tt hep-th/0112248}}.
%%CITATION = HEP-TH 0112248;%%.

\bibitem{Biller:1998hg}
S.~D. Biller {\em et al.}, ``Limits to quantum gravity effects from
  observations of tev flares in active galaxies,'' {\em Phys. Rev. Lett.} {\bf
  83} (1999) 2108--2111,
\href{http://www.arXiv.org/abs/gr-qc/9810044}{{\tt gr-qc/9810044}}.
%%CITATION = GR-QC 9810044;%%.

\bibitem{Tamaki:2002iz}
T.~Tamaki, T.~Harada, U.~Miyamoto, and T.~Torii, ``Particle velocity in
  noncommutative space-time,'' {\em Phys. Rev.} {\bf D66} (2002) 105003,
\href{http://www.arXiv.org/abs/gr-qc/0208002}{{\tt gr-qc/0208002}}.
%%CITATION = GR-QC 0208002;%%.

\end{thebibliography}\endgroup

\end{document}